\documentclass[a4paper,12pt]{article}

\usepackage[latin1]{inputenc}
\usepackage{amsmath}
\usepackage{amsfonts}
\usepackage{amssymb}
\begin{document}

\begin{titlepage}

\topmargin=3.5cm

\textwidth=13.5cm

\centerline{\Large \bf Deformation Quantization of  }

\vspace{0.4cm}

\centerline{\Large \bf  Odd Dimensional anti-de Sitter Spaces as}
\vspace{0.4cm}

 \centerline{\Large \bf   Contact Manifolds}

\vspace{1.0cm}

\centerline{\large \bf Levent Akant\footnote{E-mail:
akant@gursey.gov.tr}}

\vspace{0.5cm}

\centerline{ \textit{Feza Gursey Institute}}

\centerline{\textit{Emek Mahallesi, Rasathane Yolu No.68 }}
\centerline{\textit{Cengelkoy, Istanbul, Turkey}}

\vspace{1.5cm}

\paragraph{Abstract:} We quantize odd dimensional anti-de Sitter
spaces by applying the method of deforming contact manifolds
proposed by Rajeev in \cite{rajeev}. The construction in the present
paper consists of the identification of the odd dimensional anti-de
Sitter space as a hypersurface of contact type and the subsequent
use of 'symplectization' principle. We also show that this
construction generalizes to any odd dimensional hypersurface which
can be represented as a nonzero level set of a homogenous function.
\end{titlepage}

\baselineskip=0.6cm

\abovedisplayskip=0.4cm

\belowdisplayskip=0.4cm

\abovedisplayshortskip=0.3cm

\belowdisplayshortskip=0.3cm

\jot=0.35cm

\section{Introduction}
Recently Rajeev, in his work on quantization of thermodynamics
\cite{rajeev}, proposed a method of quantizing contact manifolds.
Basic idea behind this method is the symplectization principle of
contact manifolds. This principle allows one to set up a 1-1
correspondence between the functions on a contact manifold and a
subset $\mathcal{F}$ of functions on a symplectic manifold. Then one
uses the usual deformation of the symplectic manifold to deform
$\mathcal{F}$ and map the result back to the contact manifold. The
first order correction to the resulting star product is given by the
Legendre bracket which is the analog of Poisson bracket in contact
geometry. In \cite{rajeev} this method was applied to quantize odd
dimensional spheres.

Considering the central role of AdS/CFT correspondence \cite{mal,
gubser, wittenhol} in M-theory and the important relations between
noncommutative geometry \cite{connes} and string theory
\cite{seiwit} we find it useful to apply Rajeev's method to quantize
odd dimensional anti-de Sitter spaces.

As usual we will regard $AdS_{2n+1}$ as a hypersurface in
$\mathbf{R}^{2n+2}$ defined by the equation
\begin{equation}
    -(x^{0})^{2}-(x^{1})^{2}+(x^{2})^{2}+\ldots+(x^{2n+1})^{2}=-1.
\end{equation}
We will regard $\mathbf{R}^{2n+2}$ as a symplectic manifold with the
canonical symplectic form. Our symplectization will rely on the fact
that $AdS_{2n+1}$ is a hypersurface of contact type. Then we will
show how one can lift functions on $AdS_{2n+1}$ to a subset of
homogenous functions on $\mathbf{R}^{2n+2}$. Following the general
strategy of \cite{rajeev} we will use the star product on
$\mathbf{R}^{2n+2}$ to induce one on $AdS_{2n+1}$. The homogeneity
of the function defining $Ads_{2n+1}$ is an essential ingredient of
the constructions presented in this paper. In fact we will show that
our results can be applied to any odd dimensional hypersurface which
can be represented as a nonzero level set of a homogenous function.

\section{Contact Geometry}
In this section we will summarize basic results from contact
geometry that will be useful in our constructions. Detailed accounts
of the subject can be found in \cite{dusa, arnold}.

Let $M$ be a manifold of dimension $2n+1$ and $\xi$ a $2n$
dimensional subbundle of $TM$. We will assume that there exist a
global 1-form $\alpha$ such that $\mathrm{ker}\alpha=\xi$ and
$d\alpha$ is nondegenerate on $\xi$. Then $\xi$ is called a contact
structure and $\alpha$ is called the contact 1-form. The assumptions
on $\alpha$ are equivalent to $\mathrm{ker}\alpha=\xi$ and
$\alpha\wedge (d\alpha)^{n}\neq 0$. Since $d\alpha$ is a 2-form on
an odd dimensional manifold, it must be degenerate. However, by
assumption $d\alpha$ is nondegenerate on a $2n$ dimensional space
$\xi$. Therefore there exist a unique direction along which
$d\alpha$ is degenerate. This gives us the definition of the Reeb
vector field as the unique vector field $Y$ such that
\begin{equation}
\iota_{Y}d\alpha=0, \;\;\;\;\;\; \iota_{Y}\alpha=1.
\end{equation}
A vector field $X$ satisfying $\pounds_{X}\alpha=g_{X}\alpha$ where
$g_{X} \in C^{\infty}(M)$ is called a contact vector field. In
particular the Reeb vector field $Y$ is a contact vector field with
$g_{Y}=0$. There is a one-to-one correspondence between the set of
all contact vector fields and $C^{\infty}(M)$. Given a contact
vector field $X$ the corresponding function $H$ is defined as
$H=-\iota_{X}\alpha$. Moreover $g_{X}=-dH(Y)$. Conversely given a
function $H$ the corresponding contact vector field is $X_{H}=Z-HY$
where $Z\in \xi$ is the unique solution of $\left. \iota_{Z}d\alpha
\right \vert_{\xi}=\left.dH \right \vert_{\xi}$.

The analog of the Poisson brackets in contact geometry is the
Legendre bracket defined by
\begin{eqnarray}
  \left(F,G\right) &=& -\alpha([X_{F},X_{G}]) \\
   &=& dG(Y)F+X_{G}F \\
   &=& -dF(Y)G-X_{F}G.
\end{eqnarray}
Legendre bracket is bilinear, antisymmetric and satisfies Jacobi
identity. However the Leibnitz rule is not satisfied. Instead it
satisfies:
\begin{equation}
    (f,gh)=g(f,h)+(f,g)h+(1,f)gh
\end{equation}
which is sometimes called the generalized Leibnitz rule. Since the
Leibnitz rule is not satisfied it follows that the constant function
is not in the center of the algebra. In fact, upon quantization
\cite{rajeev} the constant function does not become a constant
multiple of the identity. This is the novel feature of quantization
on contact manifolds.

Let $Q$ be a hypersurface in a symplectic manifold $(N,\omega)$. If
there exist a 1-form $\alpha$ in $Q$ such that $d\alpha=\left.
\omega\right \vert_{Q}$ then $\alpha$ is a contact form. In this
case $Q$ is called a hypersurface of contact type. A Liouville
vector field in a symplectic manifold is defined as a vector field
satisfying $\pounds_{T}\omega=\omega$. An equivalent condition for
$Q$ to be a hypersurface of contact type is the existence of a
Liouville vector field $T$ which is transverse to $Q$.
\section{Contact Geometry of $AdS_{2n+1}$}
Consider the $2n+1$ dimensional anti-de Sitter space $AdS_{2n+1}$.
This is a hypersurface in $\mathbf{R}^{2n+2}$ defined by the
equation $R(x)=-1$ where
\begin{equation}
R(x)=-(x^{0})^{2}-(x^{1})^{2}+(x^{2})^{2}+\ldots+(x^{2n+1})^{2}.
\end{equation}
It will be useful to call the even coordinates $p$ and odd coordinates $q$. Then
\begin{equation}
R(p,q)=(-(p^{1})^{2}-(q^{1})^{2})+((p^{2})^{2}+(q^{2})^{2}+\ldots+(p^{n+1})^{2}+(q^{n+1})^{2}).
\end{equation}
We will think of $\mathbf{R}^{2n+2}$ as a symplectic manifold with
the canonical symplectic form
\begin{equation}
\omega=\sum_{k} dp^{k}\wedge dq^{k}
\end{equation}
Clearly $\omega=d\alpha_{0}$ with
\begin{equation}
\alpha_{0}=\frac{1}{2}\sum_{k}p^{k}dq^{k}-q^{k}dp^{k}.
\end{equation}
Consider the vector field $T$ in $\mathbf{R}^{2n+2}$ given by
\begin{equation}
T=\frac{1}{2}\sum_{k} p^{k}\frac{\partial}{\partial p^{k}}+q^{k}\frac{\partial}{\partial q^{k}}.
\end{equation}
It follows that $T$ is a Liouville vector field on
$\mathbf{R}^{2n+2}$. Notice that $\iota_{T}\omega=\alpha_{0}$.
Moreover we have
\begin{equation}
    \pounds_{T}R=R.
\end{equation}
This is nothing but Euler's homogenous function theorem applied to
$R$, a homogenous function of degree 2. Consequently we see that on
$AdS_{2n+1}$ where $R=-1$:
\begin{eqnarray}
\pounds_{T}R=-1\neq 0.
\end{eqnarray}
So $T$ is transversal to $AdS_{2n+1}$. Thus odd dimensional $AdS$
spaces are hypersurfaces of contact type. The contact form on
$AdS_{2n+1}$ is
\begin{equation}
\alpha=\left.\iota_{T}\omega \right \vert_{AdS}=\left. \alpha_{0}\right \vert_{AdS}
\end{equation}
Let us consider the Hamiltonian vector field corresponding to the
function $R$
\begin{equation}
\iota_{X_{R}}\omega=-dR
\end{equation}
Explicitly we have
\begin{eqnarray}
X_{R}&=&\sum_{k} -\frac{\partial H}{\partial q^{k}}\frac{\partial}{\partial p^{k}}+\frac{\partial H}{\partial p^{k}}\frac{\partial}{\partial q^{k}}\nonumber\\
&=&2\left( q^{1}\frac{\partial}{\partial p^{1}}-q^{2}\frac{\partial}{\partial p^{2}}-\ldots\right) +2\left( -p^{1}\frac{\partial}{\partial q^{1}}+p^{2}\frac{\partial}{\partial q^{2}}+\ldots\right)
\end{eqnarray}
Clearly $X_{R}$ is tangent to $AdS_{2n+1}$. For any vector field $V$
tangent to $AdS_{2n+1}$ we have
\begin{equation}
(\iota_{X_{R}}d\alpha)(V)=-dR(V)=-\pounds_{V}H=0.
\end{equation}
Thus on $AdS_{2n+1}$ we have
\begin{equation}
\iota_{X_{R}}d\alpha=0
\end{equation}
Moreover by homogeneity of $R$
\begin{equation}
\iota_{X_{R}}\alpha_{0}=\frac{1}{2}\sum_{k}p^{k}\frac{\partial
R}{\partial p^{k}}+p^{k}\frac{\partial R}{\partial p^{k}}=R
\end{equation}
which, when restricted to $AdS_{2n+1}$, gives
\begin{equation}
\iota_{X_{R}}\alpha_{0}=-1
\end{equation}
Thus $X_{R}$ is the negative of the Reeb vector field on
$AdS_{2n+1}$.
\section{Deformation Quantization of $AdS_{2n+1}$}
Now let us try to lift functions on $AdS_{2n+1}$ to
$\mathbf{R}^{2n+2}$. We will consider the functions that can be
expanded in the eigenfunctions of the Laplacian on $AdS_{2n+1}$
\cite{Isham, Breit, Bala, Filho}. Thus we can expand in e.g.
\begin{equation}
\Phi=e^{-i\nu t}Y_{l,\{m\}}(\Omega)F(\sin\rho).
\end{equation}
Here $F$ is a hypergeometric function as given in \cite{Bala},
$Y_{l,\{m\}}$ are the spherical harmonic on $S^{2n-1}$. The
coordinates of a point on $AdS_{2n+1}$ in the ambient space are
given in terms of the angular variables as
\begin{eqnarray}
p^{1}&=&\sec \rho \cos \tau\\
q^{1}&=&\sec \rho \sin \tau\\
x^{i}&=& (\tan \rho) \Omega_{i}.
\end{eqnarray}
Here $\sum_{i=1}^{n}\Omega_{i}^{2}=1$, $0\leq \rho < \frac{\pi}{2}$,
$0\leq \tau <2\pi$ and we denoted the coordinates
$p^{2},q^{2},\ldots$ by $x^{1},x^{2},\ldots$, respectively.
Expressing $\Phi$'s in terms of the original variables and
continuing the result to $\mathbf{R}^{2n+2}$ gives us a function on
$\mathbf{R}^{2n+2}$. The effect of this on the coordinates is the
substitution:
\begin{eqnarray}
  \cos^{2} \rho &\rightarrow & \frac{(p^{1})^{2}+(q^{1})^{2}-\sum_{i}(x^{i})^{2}}{(p^{1})^{2}+(q^{1})^{2}} \\
  \tau &\rightarrow & \tan^{-1}\left(\frac{q^{1}}{p^{1}}\right) \\
  \Omega_{i} &\rightarrow &
  \frac{x^{i}}{\sqrt{\sum_{i}(x^{i})^{2}}}
\end{eqnarray}
The resulting function is defined only on
$(p^{1})^{2}+(q^{1})^{2}\geq\sum_{i}(x^{i})^{2}$. Geometrically this
can be understood as follows. The transverse vector field $T$ maps a
level surface of $R$ onto another, infinitesimally close level set.
In such infinitesimal steps one covers the region in question which
is thence diffeomorphic to the trivial bundle $AdS_{2n+1}\times
\mathbf{R}^{+}T$. Notice that the resulting function is homogenous
of degree 0. In particular it is annihilated by the vector field
$T$. Given a function $F$ on $AdS_{2n+1}$ we denote its extension
described above by $F_{1}$ and define the desired lift by
\begin{equation}
    \widetilde{F}=-RF_{1}
\end{equation}
This lift is homogenous of degree $2$. Notice that
$\widetilde{1}=-R$. Now we can check that the Poisson bracket of two
such lifts gives us a function whose restriction to $AdS_{2n+1}$
gives the Legendre bracket of the original functions. Let $F$ be a
function on $AdS_{2n+1}$. We will denote the corresponding contact
vector field by $X_{F}$. The Hamiltonian vector field corresponding
to the lift $\widetilde{F}$ will be denoted by $\widetilde{X}_{F}$.
In order to compare the Poisson bracket with the Legendre bracket we
must relate Hamiltonian vectors field to contact vector fields. We
will show that the latter are the restrictions of the former on
$AdS_{2n+1}$. According to the general theory $X_{F}=Z_{F}+FX_{R}$
where $Z_{F}$ is the unique vector field in $\xi$ such that
\begin{equation}
    d\alpha(Z_{F}, V)=dF(V) \;\;\;\; \forall V\in \xi
\end{equation}
Let $\left\{V_{a}\right\}$ be a basis for $\xi$ and recall that
$d\alpha=\left.\omega \right|_{AdS}$. Expanding $Z_{F}=Z^{a}V_{a}$
we get
\begin{equation}\label{e1}
    Z^{a}\omega_{ab}=dF(V_{b}).
\end{equation}
Let us also express $\widetilde{X}_{\widetilde{F}}$ in the basis
$\left\{T,X_{H},V_{a}\right\}$ as
\begin{equation}
    \widetilde{X}_{\widetilde{F}}=A^{a}V_{a}+BT+CX_{R}.
\end{equation}
Then
\begin{eqnarray}
  \left\{\widetilde{F},\widetilde{G}\right\} &=& -\widetilde{X}_{\widetilde{F}}\widetilde{G}= \\
  &=&(\widetilde{X}_{\widetilde{F}}R)G_{1}+R\widetilde{X}_{\widetilde{F}}G_{1}
\end{eqnarray}
Now
\begin{equation}
\left.\widetilde{X}_{\widetilde{F}}R\right|_{AdS}=\left.-B\right|_{AdS}
\end{equation}
and
\begin{equation}\label{e2}
\left.\widetilde{X}_{\widetilde{F}}G_{1}\right|_{AdS}=(\left.A^{a}V_{a}G_{1}+CX_{R}G_{1})\right|_{AdS}
\end{equation}
So
\begin{eqnarray}
\left.\left\{\widetilde{F},\widetilde{G}\right\}\right|_{AdS}=\left.-B\right|_{AdS}G-\left.A^{a}\right|_{AdS}V_{a}G-\left.C\right|_{AdS}X_{R}G.
\end{eqnarray}
On the other hand we have
\begin{eqnarray}
    \iota_{\widetilde{X}_{F}}\omega&=&A^{a}\omega(V_{a},\;)+B\omega(T,\;)+C\omega(X_{R},\;)\\
    &=&A^{a}\omega(V_{a},\;)+B\alpha(\;)+C\omega(X_{R},\;)=d\left(-RF_{1}\right)=-dR F_{1}-RdF_{1}.\nonumber\\
\end{eqnarray}
Contracting this with $V_{b}$ we get
\begin{equation}
    A^{a}\omega_{ab}=-F dR(V_{b})+dF(V_{b})=dF(V_{b})
\end{equation}
Comparing this with (\ref{e1}) we see that $A^{a}=Z^{a}$. Similarly
contraction with $T$ gives
\begin{equation}
    C=F,
\end{equation}
and finally contracting by $X_{R}$ we get
\begin{equation}
    B=-dF(X_{R})=-\pounds_{X_{R}}F
\end{equation}
Substituting these into (\ref{e2}) we get the desired result
\begin{equation}
    \left.\left\{\widetilde{F},\widetilde{G}\right\}\right|_{AdS}=-dF(Y)G-X_{F}G=(F,G)
\end{equation}
Now, following Rajeev \cite{rajeev} we can define the star product
on $AdS_{2n+1}$ as
\begin{eqnarray}
  F*G &=&
  \left.\widetilde{F}\exp\left[-\frac{i\hbar}{2}\left(\frac{\overleftarrow{\partial}}{\partial q_{k}}\frac{\overrightarrow{\partial}}{\partial p_{k}}-
  \frac{\overleftarrow{\partial}}{\partial p_{k}}\frac{\overrightarrow{\partial}}{\partial
  q_{k}}\right)\right]\widetilde{G}\right|_{AdS}\\
  &=& FG-\frac{i\hbar}{2}(F,G)+\ldots.
\end{eqnarray}
This provides the deformation of $AdS_{2n+1}$ as a contact manifold.
\section{Generalizations and Conclusion}
Our results can be generalized as follows. Let $Q$ be a hypersurface
defined as a nonzero level set of a homogenous function on an even
dimensional space. If $Q$ is defined by an equation of the form
$R=a\neq 0$ where $R$ is a homogenous function of degree $r$ the
transverse Liouville vector field can be chosen as
\begin{equation}
    T=\frac{1}{r}\sum_{k} p^{k}\frac{\partial}{\partial p^{k}}+q^{k}\frac{\partial}{\partial q^{k}}.
\end{equation}
This turns $Q$ into a hypersurface of contact type. Again thanks to
the homogeneity of $R$ the Hamiltonian vector field $X_{R}$ gives
the Reeb field on the hypersurface. The crucial point in the problem
is the construction of the lift. This can be accomplished if one can
find a suitable basis for the functions on the hypersurface which
can be lifted to homogenous functions of degree $r$ in the ambient
space. In particular this construction works for odd dimensional
spheres \cite{rajeev}, for de Sitter spaces and, as we have shown
explicitly, for anti-de Sitter spaces.

\paragraph{Acknowledgement:} The author would like to thank S. G.
Rajeev and A. Kaya for useful conversations.


\begin{thebibliography}{10}

\bibitem{rajeev} S. G. Rajeev, math-ph/0703061

\bibitem{mal} J. Maldacena, Adv. Theor. Math. Phys. \textbf{2}
(1998) 231

\bibitem{gubser} S. S. Gubser, I. R. Klebanov, A. M. Polyakov, Phys.
Lett. \textbf{B428} (1998) 105

\bibitem{wittenhol} E. Witten, Adv. Theor. Math. Phys. \textbf{2}
(1998) 253

\bibitem{connes} A. Connes, \textit{Noncommutative Geometry},
Academic Press (1994)

\bibitem{seiwit} N. Seiberg, E. Witten, JHEP \textbf{9909} (1999)
032

\bibitem{dusa} D McDuff, D. Salamon, \textit{Introduction to Symplectic Topology},
Oxford Science Publications (1998)

\bibitem{arnold} V. I. Arnold, A. B. Givental, \textit{Symplectic
Geometry} in Vol. IV of \textit{Encyclopedia of Mathematical
Sciences} Eds. V. I. Arnold, S. P. Novikov, Springer-Verlag (1990)

\bibitem {Isham} S. J. Avis, C. J. Isham, D. Storey, Phys. Rev. \textbf{D18} (1978) 3564

\bibitem {Breit} P. Breitenlohner, D. Z. Freedman, Phys. Lett. \textbf{B115}
(1982) 197; Ann. Phys. \textbf{144} (1982) 249


\bibitem {Bala} V. Balasubramanian, P. Kraus, A. Lawrence Phys. Rev. \textbf{D59} 046003


\bibitem {Filho} H. Boschi-Filho, N. R. F. Braga, Phys. Lett. \textbf{B505}
(2001) 263



\end{thebibliography}
\end{document}